# Importance of pore length and geometry in the adsorption/desorption process: a molecular simulation study

M.A. Balderas Altamirano*[1, 2], S. Cordero[1], R. López-Esparza[2, 3], E. Pérez[2], A. Gama Goicochea [2, 4]

[1]Departamento de Química, Universidad Autónoma Metropolitana, Unidad Iztapalapa, D. F., Mexico.
[2]Instituto de Física, Universidad Autónoma de San Luis Potosí, San Luis Potosí, Mexico.
[3]Departamento de Física, Universidad de Sonora, Hermosillo, Sonora, Mexico.
[4]A. Schulman de México, San Luis Potosí, Mexico.
* Corresponding author. Email: nyfg@xanum.uam.mx

## Abstract

Discrete potentials can describe properly the liquid vapor boundary that is necessary to model the adsorption of gas molecules in mesoporous systems with computer simulations. Although there are some works in this subject, the simulations are still highly time - consuming. Here we show that an efficient alternative is to use the three - dimensional Ising model, which allows one to model large systems, with geometries as complex as required that accurately represent the liquid vapor boundary. In particular, we report molecular simulations of cylindrical pores of two different geometry, using a discrete potential. The effect of the length of the pore in the hysteresis loop for a finite pore and infinite one is studied in detail. Lastly, we compare our predictions with experimental results and find excellent agreement between the area of the hysteresis loop predicted for the finite pore and that found in adsorption/desorption experiments.

**Keywords**

**Gas Absorption; Mesoporous Structures; Hysteresis Loop; Ising Model**

## 1. Introduction

Mesoporous materials have become important in different areas of modern science and industry. For example, there is ample research on separation, capture and storage of gases such as $CO_2$ [1], hydrogen [2] and nitrogen [3]; their potential applications appear in electrochemical capacitors [4]; drug carriers [5, 6]; biological systems [7, 8] and nanofluidics [9] to name but a few. These applications offer opportunities for scientists to develop new strategies and techniques for the synthesis of new materials. Physical adsorption of fluids confined in narrow pores is used to characterize micro and mesoporous materials and to study their phase transitions.

The area of a surface is a fundamental property to take into account when studying the effectiveness of catalytic materials. A typical route to characterize the surface area is using gas adsorption; this technique consists of adding vapor molecules to a material, and when the pressure is increased an adsorption isotherm can be obtained. Reversing the process one gets a desorption line, and a hysteresis loop is frequently found. The hysteresis loop has been explained in terms of factors that are not yet well understood, such as meniscus formation, cavitation, and the structure of the solid on which adsorption occurs.

According to the literature, there are basically three kinds of pore structures that are capable of explaining the appearance of the hysteresis loop: independent pore, ordered and disordered pore network. Independent pores are simple structures like the cylindrical pore and the so-called ink bottle pore. The ordered pore network is a more complex system, which can be represented as a connection of independent pores, and finally the disordered pore networks are pores structures of complex nature [10].

The tools typically used to study adsorption in mesoporous structures is through the development of new materials, and more recently with computer simulations. The experimental work has increased since the findings of materials like the MCM41 [11], SBA15 [12], and SBA-16 [13], which represent pore structures similar to the cylindrical pore and the ink bottle pore. However, it has become necessary to perform also molecular simulations so that one can have total control over the variables that affect the adsorption process, for example in cylindrical pores by density functional theory [14], Monte Carlo [15] and molecular dynamics simulations [16], mostly using continuous potentials. These potentials can be an accurate representation of the adsorption process, but the computing time they require to solve the systems can be extremely long for large systems. An alternative to these methods is the three – dimensional (3D) Ising model, which is a discrete potential and therefore it requires less time to solve the motion of particles under its influence, thereby allowing one to model large pore structures with complex geometries. Some authors have applied the 3D Ising model to the study of adsorption on complex structures, such as Edison et al. [17], who considered a carbon slit and simulated the hysteresis loop adding an anisotropic point in the system. Naumov et al. [18], added roughness to a cylindrical pore, while Pasinetti et al. [19], modeled the adsorption in a one dimensional channel. The cylindrical pore, as a simple model to study the adsorption can be extended by introducing periodicity. Such extension can be used to model hysteresis loops of real systems.

There are various reports in the literature that have addressed different aspects of the adsorption – desorption phenomena in pores using models. One of those is the work of Talanquer and Oxtoby [20], who found using density functional theory (DFT), that the critical temperature of the fluid is reduced when the length of the pore is increased. Wilms and co workers [21] used Monte Carlo simulations to solve the 2D Ising model for pores of finite and infinite length and found that the critical point seen in the hysteresis loop does not correspond to the pore's critical point. In a recent review, Monson [22] discusses the modeling of the hysteresis cycle for the Ising model (lattice gas) using DFT with emphasis on two particular pore geometries, one of which is a duct pore that whose adsorption isotherm was obtained for both finite and infinite cases.

In regard to the study of the hysteresis loop in the adsorption – desorption process in finite pores, the work of Marconi and van Swol [23] is particularly relevant. They used an Ising – like Hamiltonian to model adsorption on 3D pores, and solved it using DFT within the mean

– field approximation, for two types of pore geometries. One of them was an infinitely long cylinder, with open ends. In the other case one of the ends of the cylinder was closed, to mimic pore blocking effects. They found that in the pore with a closed end, menisci appeared and disappeared at the open end of the pore, which determined the extent of the hysteresis loop. The work reported here reproduces the trends found by Marconi and van Swol by solving the Ising model for both of our 3D models of pores using Monte Carlo, which is an essentially exact method that does not rely on the mean – field approximation, taking fully into account the density fluctuations at constant chemical potential. The latter is an aspect of crucial importance for the accurate prediction of thermodynamic properties in equilibrium [24].

Given that context, in this work we report molecular simulations of finite and infinite cylindrical pores using a discrete, 3D Ising potential, and compare our predictions with experimental adsorption isotherms. This article is organized as follows: In the section 2 we describe the methodology, also, the details of the cylindrical solid are presented. In section 3, we report the isotherms of two kinds of pores, finite and infinite at different lengths and compare the results with experimental isotherms. The conclusion can found in section 4.

## 2. Methodology

We performed Monte Carlo simulations using the Grand Canonical ensemble (constant volume, $V$, temperature, $T$, and chemical potential of the bulk fluid, $\mu$) to describe the adsorption process in both types of cylinders. In the simulations we have a cage with positions, each position represent a solid or a position available to be filled with fluid. This model assumes that the kinetic energy of the gas particles is constant and one only has to calculate the potential energy. Initially, we define the solid structure and then determine if a given site can be occupied by a fluid particle or not using the Metropolis algorithm [15] to accept or reject a molecule, in Monte Carlo simulations.

The discrete potential used in these simulations is [25]:

$$H = -J\sum n_i n_j t_i t_j - \mu \sum n_i t_i - yJ \sum \left[ n_i t_i (1-t_j) + n_j t_j (1-t_i) \right]$$ (1)

where $n_i = 0, 1$ is the fluid occupation variable, and $t_i$ is the solid occupancy variable, representing whether the site $i$ is occupied by the solid ($t_i = 0$) or available for the fluid ($t_i = 1$). The first sum takes into account the interaction between the fluid particles, $J$ is the fluid-fluid interaction; the second sum counts the $\mu$ (chemical potential) contribution to the system. The third term takes into account the attractive interactions between the fluid particles and the solid surface; the parameter $y$ rescales the $J$ value to represent different attractions in the solid pore. The double summations run over all of the distinct nearest neighbor pairs. We fixed the width of the solid walls so that their thickness is equal to the size of two molecules. In the filling and emptying processes, we chose $\mu$ so that the resulting adsorption of molecules is near zero, then it is increased gradually at $\Delta\mu$=0.01 intervals. At each $\mu$ one obtains a configuration, which is then used for the next value of $\mu$, until a full adsorption isotherm is obtained. The program stops when all the fluid positions inside the cylindrical pore are filled with fluid. Then the $\mu$ potential is gradually reduced at the same rate, until the pore is empty. The number of attempts to insert or remove particles was $10^4 \times N_{total}$ where $N_{total}$ is the number of molecules in the system. Throughout this work reduced units are used: $U^*=U/kT_c$, $\rho^* = N/N_{total}$, and $\mu^* = \mu/kT_c$. $T_c$ is the critical temperature which corresponds to 127 K for $N_2$. *N* is the average number of molecules accepted in the simulation, $\rho^*$ is the relative density of the fluid. The *J* variable is chosen equal to 0.7447 [26, 27], and the parameter *y* was taken as 2.0 [28, 29].

The two structures used in our simulations are shown in *Fig. 1*. The difference between these two structures are the types of ends of the pores. In the infinite pore, a), we have periodic boundary conditions (PBC), while in the finite pore, b) there is a portion of space added to simulate the vapor - liquid interface, but PBC are also enforced at the ends of the simulation box. Several lengths of the pores were modeled: 180, 108, 36, 12 and 2.4 nm. The radius R in all cases was 2.1 nm. The area available for adsorption was equal for both pores shown in *Fig. 1*, to make comparisons.

## 3. Results and discussion

In the literature one finds many reports where the radius of the pore is varied, but there are no studies to the best of our knowledge on the effect of the length of the pore during adsorption and desorption processes, which are crucial for the understanding of the hysteresis loop. We start by reporting the effect of cylinder length in finite and infinite cylindrical pores.

**3.1 Influence of cylinder length and pore geometry on the hysteresis loop**

To characterize the effect of the length of the cylindrical pore in the adsorption isotherms, we constructed cylindrical pores with a radius of 2.1 nm in all the cases. The infinite pore simulations are reported in *Fig. 2*. All the isotherms obtained have a "knee" in their relative density at the same value of relative pressure: P/Po~0.25, which signals the monolayer formation. We see also in *Fig. 2*, that all the adsorption isotherms fall on the same curve, except at the 2.4 nm of pore length. Notice also that if the length of the mesopore is increased, the width of the hysteresis loop is reduced.

In *Fig. 3* we show the adsorption/desorption isotherms for a finite cylindrical pore, which has a vapor gap at the ends of the pore. As the pore length is increased, the width of the hysteresis loop increases also, in sharp contrast with the trend found for the infinite pore (see *Fig. 2)*. The adsorption isotherm is almost the same in all cases, except for the pore with the smallest length (12 nm), as occurs also for the infinite pore. Additionally, there appears the "knee" signaling the monolayer formation at relative pressure equal to 0.25, as found in *Fig. 3*. Although the adsorption isotherms fall more or less on the same curve, in the reversed process there is a slight difference between the desorption isotherms, as shown in the inset in *Fig. 3*. If the length of the pore is increased, the width hysteresis loop increases too; this means that more energy is required to desorb fluid molecules from the surface.

The results shown in *Figs. 2* and *3* indicate that when the pore length is increased, more molecules are adsorbed, which in turn increases the energy necessary to desorb particles from the fluid in the reverse process; this is the reason for the hysteresis loops in both figures. The differences between the hysteresis loops in the finite pore structures are rather poor, as shown in *Fig. 3*. The inset in that figure shows just how small these differences are. The space gaps at both ends of the finite cylindrical pore (see *Fig. 1b*) make up a region for vapor – liquid coexistence, which makes it energetically more favorable to desorb particles that for the infinite pore case.

As is depicted in *Fig. 4*, the width of the hysteresis loop is considerably larger for the infinite pore than that in the finite pore. This difference can be understood in terms of the vapor interface that forms at the ends of the finite pore, while in the infinite pore there is only fluid, hence one expects a greater hysteresis loop primarily because of the condensate formation. By contrast, in the finite case there are interfaces at the ends of the cylinder, so the sorbate molecules can interact with vapor molecules at such interfaces, which requires less energy for desorption, thereby reducing the total free energy of the system. Also, one must bear in mind that actual pores are never infinite in the sense of *Fig. 1(a)* and resemble more the geometry in *Fig. 1(b)*.

### 3.2 Comparison with experimental results

To test the usefulness of the 3D Ising model to reproduce experimental adsorption/desorption isotherms, we have compared our results with those reported by Ojeda et al. [30] for the $N_2$ adsorption in silica mesopore systems, at radii from 4 nm up to 6 nm. We computed the area of the hysteresis loops in *Fig. 4* and also those reported in [30]; these results are listed in Table I. The agreement between the hysteresis area predicted for the finite pore and the experimental results is remarkable, especially given the simple nature of our model and geometry of the pores, while the experiments were performed on highly complex structures. By contrast, the area of the hysteresis loop found for the infinite pore is more than 400 % larger than the one found in the experiments of Ojeda et al. [30, 31]. Therefore, one must be careful when using overly simplified geometries in models of pores solved through computer simulations. The comparison of isotherms that complements this Table and *Fig*. 4 can be found in [31].

Lastly, there is also something to be said about the effects of finite size, which are unavoidable in any computer simulation. As seen in *Figs. 2 and 3*, the adsorption isotherms for both infinite and finite pores of the smallest length (2.4 nm and 12 nm, respectively) are somewhat larger than those for larger pores. However, as the length of the pores is increased the isotherms tend to fall on the same curve, meaning that the limiting behavior has been reached and that finite size effects are minimal. The CPU time required to perform simulations with these larger pores can be more than compensated with the computational efficiency of the Ising model.

**4. Conclusions**

In this work we report Monte Carlo simulations of finite and infinite cylindrical pores at various lengths, to obtain adsorption/desorption isotherms, using the 3D Ising model. The difference between those types of pores is the absence or presence of a slice of space at the ends of the cylinder. In the case of the so called infinite pore, PBC were imposed directly on the faces perpendicular to the axis of the cylinder, hence its name. For the finite pore a space with vapor was added at each end, and PBC were applied. The reason for modeling both types lies on the fact that while most simulations on this topic use the infinite cylindrical pore, we argue that pores found in nature can be more appropriately and simply modeled with a finite pore. We have explored the consequences that those geometrical shapes have on measurable properties, such as adsorption and desorption of simple gases. The results show that while the trends are qualitatively similar, there are important differences between the isotherms of infinite and finite pores, particularly in regard to the width of the hysteresis loop, being smaller for the latter ones. It is argued that this phenomenon is a consequence of the existence of a vapor phase in the spaces added at the ends of the finite cylindrical pore, where it is energetically less costly to remove gas particles when the chemical potential is reduced, to yield the desorption curve.

Our predictions were compared with experimental adsorption/desorption curves taken from the literature, and good agreement was found between the area of the hysteresis loop predicted for the finite pore and the experiments, while the area of the hysteresis loop of the infinite pore turned out to be more than 400% larger than in experiments. A finite size effect is found for the smallest length in each pore, which disappears as this length is increased. For the case of the infinite pore we find that as the length of the pore is increased the width of the hysteresis loop is reduced, while the opposite trend is found for the finite pore. These are also a consequence of finite size, becoming less pronounced as the length of the pore and the size of the simulation box are increased. We expect this work to be useful for research on more complex geometries that include the region of vapor − liquid coexistence, with more sophisticated interaction models.

## Acknowledgements

MABA thanks COMECYT, México for financial support. MABA, AGG and RLE thank IFUASLP for its hospitality and Universidad de Sonora (UNISON) for permission to run simulations in their *Ocotillo* cluster. RLE thanks UNISON for sabbatical support.

1503.04384.